# Photostriction-tunable Polarization and Structural Dynamics in Interlayer Sliding Ferroelectrics

Kun Yang[1,2,#], Jianxin Yu[1,2,#], Jia Zhang[3], Sheng Meng[4,5], Jin Zhang[1,*]

[1]*Laboratory of Theoretical and Computational Nanoscience, National Center for Nanoscience and Technology, Chinese Academy of Sciences. Beijing 100190, China*

[2]*University of Chinese Academy of Sciences, Beijing 100049, China*

[3]*Max Born Institut für Nichtlineare Optik und Kurzzeitspektroskopie, Berlin 12489, Germany*

[4]*Beijing National Laboratory for Condensed Matter Physics, and Institute of Physics, Chinese Academy of Sciences, Beijing 100190, China*

[5]*Songshan Lake Materials Laboratory, Dongguan, Guangdong 523808, China*

[#]Kun Yang and Jianxin Yu contributed equally to the work.

*Corresponding author: Jin Zhang (jinzhang@nanoctr.cn)

Two-dimensional ferroelectrics with robust polarization offer promising opportunities for non-volatile memory, field-effect transistors, and optoelectronic devices. However, the impact of lattice deformation on polarization and photoinduced structural response remains poorly understood. Here, we employ first-principles calculations to demonstrate photodoping-induced lattice expansion in rhombohedrally stacked bilayer $MoS_2$, revealing a strong coupling between photodoping carrier and lattice structure. We identify a pronounced photostrictive response in sliding ferroelectrics, wherein electron-hole excitation leads to substantial in-plane expansion, increased interlayer spacing, and enhanced ferroelectric polarization. This strain-induced modulation drives significant bandgap renormalization. The photostriction-tunable polarization and structural dynamics arise from the strong electromechanical coupling inherent to the non-centrosymmetric rhombohedral stacking. The findings provide critical insights into the nonthermal lattice expansion governing sliding ferroelectrics at atomic-scale timescales, while simultaneously laying the groundwork for next-generation electronic and memory technologies by leveraging lattice-tunable polarization switching.

Sliding ferroelectricity represents a class of polarization phenomena in layered van der Waals (vdW) materials, originating from the spontaneous breaking of interlayer translational symmetry [1-9]. In contrast to conventional ferroelectricity, which relies on ionic displacements [10-12], sliding ferroelectricity is driven by relative lateral shifts between adjacent layers, enabling reversible polarization switching without disrupting intralayer bonding. This mechanism has been experimentally realized in systems such as rhombohedral molybdenum disulfide (3R-$MoS_2$) [1,13,14] and boron nitride (rBN) [15,16], where controlled sliding transforms stacking orders and induces out-of-plane polarization. Sliding ferroelectrics offer an optomechanical platform for non-volatile memory and electronic systems, with switching feasible via lateral interlayer sliding. Experimentally, Bian and coauthors demonstrated fatigue-resistant ferroelectricity achieved through interlayer sliding in bilayer 3R-$MoS_2$ [1]. Moreover, interlayer sliding and symmetry breaking have emerged as alternative pathways to slidetronics and piezoelectrics, without the need for twist angles or moiré patterns [16,17].

The photostriction effect, light-induced lattice deformation (expansion or contraction) mediated by optical excitation, enables ultrafast control of lattice structures and electronic properties [18-23]. Harnessing the effect in sliding ferroelectrics in vdW systems paves the way for light-induced change of polarization and stacking-dependent electronic structure. This coupling not only permits reversible and localized control over ferroelectric states but also allows precise tuning of band structures, with potential applications in broadband photodetectors and strain-tunable electronic components. Besides, optical excitation can trigger a complete polarization reversal in sliding ferroelectrics [21]. Gao and coauthors revealed that photoexcitation robustly tunes the out-of-plane polarization in 3R-$MoS_2$ through a combination of photoinduced structural distortions and charge redistribution [23]. Through ultrafast surface-sensitive microscopy and in situ atomic-scale imaging, Wang *et al.* established that photoinduced thermal lattice expansion by 1.5% in monolayer $WS_2$ accelerates carrier diffusion, linking structure dynamics to optoelectronic performance [24]. While prior studies have proved the significant influence of optical excitation on both structural and carrier dynamics, the microscopic mechanisms governing the interplay between photoinduced lattice expansion and interlayer sliding in two-dimensional ferroelectrics still remain elusive, raising fundamental challenges in the atomic-scale optical control of ferroelectric order.

In this Letter, we investigate the coupling between photostriction and interlayer sliding in vdW ferroelectrics, uncovering the conditions that optical excitation can drive relative layer displacements and dynamically modulate interlayer polarizations. We find that the optomechanical interplay plays a pivotal role in light-tunable ferroelectric functionality, with a particular focus on bilayer 3R-MoS$_2$. First-principles calculations reveal that lattice expansion reduces the bandgap of 3R-MoS$_2$ and introduces structural deformation, significantly highlighting a strong light-matter interaction. Such lattice tunability is attractive for low-power electronics, where promoted carrier injection and reduced contact resistance are critical. When combined with intrinsic polarization, this supports the development of novel devices such as ferroelectric field-effect transistors with optically programmable and non-volatile switching behavior. Furthermore, the adjustable band-edge steepness offers additional control over carrier dynamics and nonlinear optical responses, which holds great promise for optoelectronic devices in low-dimensional ferroelectric systems.

The atomic and electronic structures of bilayer 3R-MoS$_2$ are illustrated in Fig. 1, with in-plane lattice vectors shown in blue. The geometric structure features a rhombohedral stacking order, characterized by a relative in-plane shift between adjacent MoS$_2$ layers (Fig. 1a). The charge density difference (Fig. 1b) is computed as the difference between the charge density of the bilayer and the superposition of isolated monolayers, revealing the interlayer charge redistribution arising from the broken inversion symmetry during the sliding-induced ferroelectric transition. The interlayer orbital hybridization generates localized dipole moments from S (lower layer) to Mo (upper layer), indicates a net upward polarization. In addition, the band structure exhibits an indirect semiconducting character with a bandgap of 1.29 eV, as shown in Fig. 1c.

We next perform real-time time-dependent density functional theory simulations (TDDFT) [25-30] to explore laser-induced carrier dynamics in 2D ferroelectric systems, revealing intricate quantum interactions between the lattice and photoexcited carriers. To monitor the carrier doping levels triggered by laser excitation in 2D ferroelectrics, Gaussian-envelope profiles are employed to define the laser pulses: $E(t) = E_0 \cos(\omega t) \exp\left[-\frac{(t-t_0)^2}{2\sigma^2}\right]$, where E$_0$, ω, t$_0$, and σ are the maximum strength, the photon energy, the peak time of the electric field, and the width of the Gaussian pulse, respectively. In response to optical stimulation, the electronic subsystem experiences direct excitation, initiating structural evolution through effective electron-phonon coupling.

In our TDDFT calculations, laser pulses with different intensities are applied to stimulate the system with the photon energy of 2 eV, which is above the bandgap of bilayer 3R-MoS$_2$ (1.29 eV on the level of PBE functional) [31-33]. Photoexcited carriers are generated by forming electron-hole pairs under optical illumination. Laser-induced carrier density ($n_{ph}$) is defined as the number of electrons excited from valence bands to conduction bands at the end of the laser pulse. The density of excited electrons is calculated by projecting the time-evolved wavefunctions on the basis of the ground state. For the electric field intensity $E_0 = 0.64 \times 10^{-4}$ V/Å, we obtain a low doping carrier density of 0.1 e/unit cell (u.c.). By contrast, the total density of excited carriers grows to 0.2 e/u.c. for $E_0 = 1.25 \times 10^{-4}$ V/Å. Further rising $E_0$ to $2.50 \times 10^{-4}$ V/Å boosts the carrier density to higher doping level of 0.8 e/u.c. Therefore, the densities of excited carriers under laser pulses rises proportionally to the laser field strengths and stabilize after photoexcitation.

Building on this understanding, we extend our analysis to the impact of photodoping on the lattice constants of the 2D sliding ferroelectric system, specifically focusing on the structural response of 3R-MoS$_2$ under optical excitation. The redistribution of charge carriers alters the potential energy landscape of the atomic structures after photo absorption, inducing internal stresses that manifest as structural deformation. As illustrated in Fig. 2a, an equilibrium lattice constant of 3.15 Å is obtained for unperturbed 3R-MoS$_2$, in good agreement with the experimental value [34], thereby validating the reliability of our computational framework. Following photoexcitation with a doping level of $n_{ph}$ = 0.1 e/u.c., we observe a discernible expansion of approximately 1%, with the in-plane lattice constant increasing to 3.18 Å (Fig. 2b). This corroborates a direct coupling between the photogenerated carrier densities and the atomic structure. When the doping level is further grown to $n_{ph}$ = 0.2 e/u.c., the lattice displays a larger expansion of 3.21 Å, corresponding to an overall expansion of nearly 2% relative to the equilibrium state (Fig. 2c). Under a photodoping level of 0.8 e/u.c., the system illustrates markedly altered energy landscape compared to the equilibrium condition, as shown in Fig. 2d. The total energy minimum occurs at a lattice constant of 3.30 Å, representing a substantial expansion, which reflects strong electron-lattice coupling under high carrier densities. The expansion under optical illumination underscores the sensitivity of the sliding ferroelectric framework to photoinduced charge injection. The observed expansion is attributed to a pronounced photostriction effect, wherein the redistribution of electronic densities following photoexcitation leads to modifications in interatomic forces and bond lengths.

These results compellingly show that photodoping not only tunes the electronic structure but also serves as a powerful tool for lattice modulation. Notably, the laser-induced structural relaxation observed in 3R-$MoS_2$ may alleviate residual strain typically present in as-grown samples, improving the overall structural integrity. This process significantly enhances the thermal and operational stability of 3R-$MoS_2$ in functional devices, contributing to the long-term reliability of optoelectronic systems. The ability to optically engineer lattice and electronic properties in real time also positions photodoping as a promising technique for the development of highly responsive, tunable materials for flexible electronics, photodetectors, and energy-efficient devices. Besides, the application of large strain can also give rise to emergent properties, such as topologically nontrivial states or phase transitions, depending on the lattice expansion magnitude or doping conditions. This offers different prospects for structure engineering in 2D sliding ferroelectrics, allowing precise control over both mechanical and electronic properties.

Different from conventional ferroelectrics, where polarization originates from ionic displacements, sliding ferroelectricity in bilayer 3R-$MoS_2$ arises from out-of-plane polarization facilitated by lateral shifts between adjacent layers. This feature facilitates reversible polarization switching without altering intralayer bonding. Experimentally, the ferroelectric states can be controlled by applying lateral stress, introducing strain through substrate interactions, or modulating the electrostatic environment via gate electrodes. These external perturbations result in relative interlayer translation, effectively switching the polarization direction. High-resolution imaging and electrical measurements have confirmed the stability and reversibility of these polarization states under optical excitation. Theoretically, first-principles calculations have shown the energy landscape governing interlayer sliding and polarization. These studies indicate that lattice strain, carrier doping, and electric fields can considerably lower the energy barriers between degenerate polarization states, offering practical routes for tuning ferroelectric behaviors.

Recent theoretical developments have further demonstrated the impact of photoexcitation in modifying interlayer interactions, underscoring that optical control of polarization may be achievable through light-induced lattice expansion [23-24]. According to our simulations, photoexcitation induces in-plane lattice expansion, which inevitably disturb the microstructure of 3R-$MoS_2$. To gain a deeper understanding of the microscopic mechanisms behind the photoinduced expansion and its impact on layer ferroelectric materials, we systematically analyze the fundamental relationship between expansive strain and interlayer characteristics including spacing, polarization and vibration.

Figure 3(a) displays the dependence of the interlayer distance on in-plane lattice expansion in 3R-$MoS_2$. As the lattice is incrementally expanded, a linear dependence of the interlayer spacing is observed, pointing to a strong coupling between in-plane and out-of-plane lattice degrees of freedom. The fully relaxed interlayer distance is 3.06 Å for the lattice without any expansion. Regarding a 5% lattice expansion, the interlayer distance decreases to 2.95 Å. This response is a hallmark of photostrictive behavior, wherein photoexcitation leads to lattice expansion via nonthermal carrier redistribution or exciton formation, resulting in spontaneous strain. Notably, the weak vdW interaction between adjacent layers in 3R-$MoS_2$ renders the system particularly susceptible to the out-of-plane relaxations, even under relatively modest in-plane strain.

Furthermore, Fig. 3(b) quantifies the variation in out-of-plane polarization as a function of lattice expansion. The polarization is found to grow significantly with the tensile strain, evidencing an emergent electromechanical response. The polarization increases monotonically from 0.84 pC/m at zero expansion to 0.89 pC/m, 1.00 pC/m, and 1.09 pC/m under 1%, 3%, and 5% expansion, respectively. This enhancement is attributed to the strain-induced interlayer charge redistribution form modified atomic displacements within the lattice expansion, which collectively strengthen the dipolar response of the system. The observed expansion-polarization relationship underscores the role of lattice strains in tailoring the dielectric and ferroelectric properties of layered transition metal dichalcogenides.

Given the observed dependence of polarization and expansion, it is essential to know how stacking orders work on this effect. In view of this, we compare the structural and electronic responses of 3R- and 2H-stacked MoS$_2$ under in-plane expansions (see Fig. S2). Upon the biaxial strain, the interlayer distance in 3R-stacked MoS$_2$ decreases significantly from 3.04 Å to 2.94 Å, whereas the 2H-stacked counterpart exhibits only a slight reduction from 3.05 Å to 3.02 Å. This difference is a direct consequence of enhanced interlayer interplay assisted by out-of-plane polarization in the 3R phase, which facilitates greater out-of-plane relaxation under tensile strain. In contrast, the centrosymmetric 2H stacking lacks intrinsic polarization, thus shows minimal changes of interlayer spacing as lattice expands. Therefore, compared to 2H phase, 3R-MoS$_2$ displays markedly stronger structural flexibility under in-plane lattice expansion. The bandgap tunability in the 3R phase stems from the structural flexibility and stronger adjustment of electronic interactions, which are more lattice-sensitive.

To quantify the prediction, we present the evolution of the electronic bandgap with lattice expansions in Fig. 3(c) and Fig. S2. A marked and progressive reduction in the bandgap is observed. The indirect bandgap of bilayer 3R phase decreases from 1.30 eV for the unexpanded lattice to 0.22 eV under 5% expansion while the 2H phase shows a relatively small decrease from 1.29 eV to 0.37 eV, realigning the pronounced sensitivity of the electronic structure to strain-induced atomic rearrangements. The bandgap narrowing under photostriction opens the possibility of optical bandgap engineering via light-induced lattice deformation, offering an approach toward tunable electronic and excitonic functionality. Collectively, the finding highlights how photostrictive strain in 3R-MoS$_2$ enables simultaneous control over structural, polarization, and electronic degrees of freedom, pointing toward slidetronics and ferroelectronics. It should be noted that the strain-induced shift of the bandgap from the visible into the near-infrared regime markedly expands the optoelectronic application of 3R-MoS2. These features collectively suggest a new class of lattice programmable optoelectronic devices that operate beyond the capabilities of conventional 2D semiconductors.

To further elucidate the influence of lattice expansion on the ferroelectric polarization switching mechanism, we perform nudged elastic band calculations [35,36] to determine the energy barrier along the minimum-energy switching pathway under varying degrees of in-plane expansion. Our results reveal that both the sliding distance and the energy barrier increase linearly as the lattice expands. As depicted in Fig. 3(d), when the lattice is expanded from 0% to 5%, the energy barrier along the low-energy pathway rises modestly from 18 to 22 meV/u.c.. These relatively low barriers indicate that the switching process remains energetically accessible, and that moderate lattice expansion exerts only a strong influence on the switching. The gradual elongation in oscillation period under tensile strain reflects a clear phonon softening effect induced by the expansion. As the lattice stretches, the reduced interatomic force constants lower the vibrational frequencies, highlighting the weakening of lattice rigidity. The robustness against structural strains reflects the intrinsic stability and potential controllability of ferroelectric switching.

Subsequently, we explore the lattice vibrations in 3R-$MoS_2$ under varying expansions. The initial phonon mode is prepared by setting the suppressed interlayer distance along perpendicular direction to excite the interlayer lattice vibrations. We then propagate the electron-phonon coupled dynamics and analyze the interlayer distances after excitations within the TDDFT scheme [25-29]. Fig. 4(a) shows the time-evolution of interlayer distances, with an oscillation of ~1.5 Å without lattice expansion. Especially, the coupled electron-ion response exhibits periodic fluctuations with a time scale of 76.0 fs. The vibration is attributed to the $A_{1g}$ mode of the ferroelectric phase [37,38]. In bilayer 3R-$MoS_2$, the $A_{1g}$ mode is a low-frequency phonon mode characterized by out-of-plane oscillations of adjacent layers, serving as a sensitive indicator of interlayer coupling strength and vertical compressibility. Due to the weak vdW interaction and the absence of inversion symmetry in 3R stacking, this mode is particularly susceptible to external perturbations [37-40].

The periodic fluctuation becomes increasingly pronounced with enhanced lattice expansion. Specifically, at 1% expansion, the fluctuation reaches 80.6 fs, indicating a slight deviation from the unstrained behavior, which is also evidenced by the progressive increasing fluctuation of 82.3 fs at 3% lattice expansion to 86.6 fs at 5% expansion. (corresponding to 385 $cm^{-1}$). This enhancement reflects the expansion-induced softening of the $A_{1g}$ phonon, indicating an increased anharmonicity and reduced restoring force between layers. Furthermore, we analyze the lattice expansion dependent phonon frequency of the interlayer $A_{1g}$ mode, as shown in Fig. 4(b). The frequency exhibits a clear linear decrease from 405 $cm^{-1}$ in the pristine lattice to about 369 $cm^{-1}$ under 5% expansion, which is in good agreement with the real-time propagation. The elongation of the oscillation period under tensile strain suggests a modified interlayer interplay. Moreover, the larger oscillation amplitude underlines the enhanced out-of-plane polarizability, which can further modulate the ferroelectric switching. These findings emphasize the tunability of coupled electron-phonon interactions in sliding ferroelectrics, offering a promising route to engineer ultrafast switching and energy dissipation pathways in layered sliding ferroelectrics.

Our real-time TDDFT calculations show that under in-plane tensile strain, the lattice vibration frequency exhibits a significant shift, signaling a softening of the interlayer restoring force. This response arises from the photo-induced in-plane lattice expansion, which indirectly weakens interlayer cohesion and facilitates greater layer separation. Simultaneously, the fluctuation of the total polarization associated with interlayer displacement is enlarged, reflecting the lattice asymmetry and flexural freedom in the distorted stacking geometry. These results indicate that strain can effectively alter the structure dynamics while amplifying the electromechanical coupling intrinsic to sliding ferroelectricity. Such tunable phonon responses present opportunities for engineering lattice-sensitive functionalities in vdW layered systems.

The combination of photostriction enhanced ferroelectricity and structural dynamics in 3R-$MoS_2$ opens up exciting possibilities for multifunctional devices that seamlessly integrate memory, and sensing capabilities. The ability to reversibly modulate both polarization and electronic structure via lattice expansion or optical fields supports precise, programmable control over key device parameters such as conductivity, threshold voltage, and optical absorption. The flexibility is particularly advantageous for non-volatile memory applications, where ferroelectric switching allows for the encoding of binary states, while the additional capability to change the bandgap introduces an analog degree of freedom. The finding paves the way for multi-level memory systems, where more than two different states are accessible for neuromorphic computing. The integration of these functionalities within a single vdW material offers significant advantages in terms of compactness, energy efficiency, and high functional density, making it ideal for the next-generation adaptive electronics, where performance and versatility are paramount.

In conclusion, combing first-principles and real-time TDDFT simulations, we have demonstrated a pronounced photostrictive effect induced by photodoping in sliding ferroelectric bilayer 3R-MoS$_2$, manifested as significant in-plane lattice expansion, modified interlayer spacing, and enhanced out-of-plane polarization. This nonthermal lattice response originates from the strong coupling between photoexcited carriers and the broken inversion symmetry of the rhombohedral stacking, enabling dynamic modulation of ferroelectric polarization and band structure on ultrafast timescales. The photostriction-driven lattice deformation and structural dynamics lead to substantial bandgap renormalization and a tunable energy barrier for polarization switching, while preserving energetically accessible switching pathways under moderate strain. Our findings establish a new optomechanical paradigm for controlling ferroic order in two-dimensional vdW materials, opening pathways for next-generation multifunctional devices such as optically programmable ferroelectric memories, strain-tunable electronics, and ultrafast optoelectronics. This work thus lays the theoretical foundation for photodoping-engineered "slidetronics" and highlights the promise of light-induced structural dynamics as a versatile tool for manipulating low-dimensional ferroelectrics.

## Acknowledgments

This work was supported by the starting funding from National Center for Nanoscience and Technology. This work was supported by the National Key R&D Program of China (2022YFA1203200), the Basic Science Center Project of the National Natural Science Foundation of China (22388101), the Strategic Priority Research Program of the Chinese Academy of Sciences (XDB36000000), the National Natural Science Foundation of China (12125202).

## Author contributions

J.Z. designed the research. All authors contributed to the analysis and discussion of the data and the writing of the manuscript.

**Conflict of Interest:** The authors declare no competing financial interest.

## Reference

[1] R. Bian *et al.*, Developing fatigue-resistant ferroelectrics using interlayer sliding switching, Science **385**, 57 (2024).

[2]  P. Meng *et al.*, Sliding induced multiple polarization states in two-dimensional ferroelectrics, Nat. Commun. **13**, 7696 (2022).

[3]  L.-P. Miao, N. Ding, N. Wang, C. Shi, H.-Y. Ye, L. Li, Y.-F. Yao, S. Dong, and Y. Zhang, Direct observation of geometric and sliding ferroelectricity in an amphidynamic crystal, Nat. Mater. **21**, 1158 (2022).

[4]  B. Qin *et al.*, Interfacial epitaxy of multilayer rhombohedral transition-metal dichalcogenide single crystals, Science **385**, 99 (2024).

[5]  F. Sui *et al.*, Atomic-level polarization reversal in sliding ferroelectric semiconductors, Nat. Commun. **15**, 3799 (2024).

[6]  P. Tang and G. E. W. Bauer, Sliding Phase Transition in Ferroelectric van der Waals Bilayers, Phys. Rev. Lett. **130**, 176801 (2023).

[7]  M. Vizner Stern, S. Salleh Atri, and M. Ben Shalom, Sliding van der Waals polytypes, Nat. Rev. Phys. **7**, 50 (2025).

[8]  M. Vizner Stern *et al.*, Interfacial ferroelectricity by van der Waals sliding, Science **372**, 1462 (2021).

[9]  K. Yasuda *et al.*, Ultrafast high-endurance memory based on sliding ferroelectrics, Science **385**, 53 (2024).

[10] S. Assavachin and F. E. Osterloh, Ferroelectric Polarization in $BaTiO_3$ Nanocrystals Controls Photoelectrochemical Water Oxidation and Photocatalytic Hydrogen Evolution, J. Am. Chem. Soc. **145**, 18825 (2023).

[11] J. F. Scott, Applications of Modern Ferroelectrics, Science **315**, 954 (2007).

[12] Y. Jiang *et al.*, Enabling ultra-low-voltage switching in $BaTiO_3$, Nat. Mater. **21**, 779 (2022).

[13] D. Yang, J. Liang, J. Wu, Y. Xiao, J. I. Dadap, K. Watanabe, T. Taniguchi, and Z. Ye, Non-volatile electrical polarization switching via domain wall release in $3R-MoS_2$ bilayer, Nat. Commun. **15**, 1389 (2024).

[14] T. H. Yang *et al.*, Ferroelectric transistors based on shear-transformation-mediated rhombohedral-stacked molybdenum disulfide, Nat. Electron. **7**, 29 (2024).

[15] L. Wang *et al.*, Bevel-edge epitaxy of ferroelectric rhombohedral boron nitride single crystal, Nature **629**, 74 (2024).

[16] K. Yasuda, X. Wang, K. Watanabe, T. Taniguchi, and P. Jarillo-Herrero, Stacking-engineered ferroelectricity in bilayer boron nitride, Science **372**, 1458 (2021).

[17] L. Rogée, L. Wang, Y. Zhang, S. Cai, P. Wang, M. Chhowalla, W. Ji, and S. P. Lau, Ferroelectricity in untwisted heterobilayers of transition metal dichalcogenides, Science **376**, 973 (2022).

[18] C. Paillard, B. Xu, B. Dkhil, G. Geneste, and L. Bellaiche, Photostriction in Ferroelectrics from Density Functional Theory, Phys. Rev. Lett. **116**, 247401 (2016).

[19] Z. Xiang, Y. Chen, Y. Quan, and B. Liao, High-Throughput Search for Photostrictive Materials Based on a Thermodynamic Descriptor, J. Am. Chem. Soc. **146**, 33732 (2024).

[20] H. Tsai *et al.*, Light-induced lattice expansion leads to high-efficiency perovskite solar cells, Science **360**, 67 (2018).

[21] Q. Yang and S. Meng, Light-Induced Complete Reversal of Ferroelectric Polarization in Sliding Ferroelectrics, Phys. Rev. Lett. **133**, 136902 (2024).

[22] J. Wang, X. Li, X. Ma, L. Chen, J.-M. Liu, C.-G. Duan, J. Íñiguez-González, D. Wu, and Y. Yang, Ultrafast Switching of Sliding Polarization and Dynamical Magnetic Field in van der Waals Bilayers Induced by Light, Phys. Rev. Lett. **133**, 126801 (2024).

[23] L. Gao and L. Bellaiche, Large Photoinduced Tuning of Ferroelectricity in Sliding Ferroelectrics, Phys. Rev. Lett. **133**, 196801 (2024).

[24] L. Wang *et al.*, Lattice Expansion Enables Large Surface Carrier Diffusion in WS2 Monolayer, ACS Energy Lett. **10**, 1741 (2025).

[25] E. Runge and E. K. U. Gross, Density-Functional Theory for Time-Dependent Systems, Phys. Rev. Lett. **52**, 997 (1984).

[26] C. Lian, M. Guan, S. Hu, J. Zhang, and S. Meng, Photoexcitation in Solids: First-Principles Quantum Simulations by Real-Time TDDFT, Adv. Theory Simul. **1**, 1800055 (2018).

[27] D. Sánchez-Portal, P. Ordejón, E. Artacho, and J. M. Soler, Density-functional method for very large systems with LCAO basis sets, Int. J. Quantum Chem. **65**, 453 (1997).

[28] M. S. José, A. Emilio, D. G. Julian, G. Alberto, J. Javier, O. Pablo, and S.-P. Daniel, The SIESTA method for ab initio order-N materials simulation, J. Phys.: Condens. Matter **14**, 2745 (2002).

[29] S. Meng and E. Kaxiras, Real-time, local basis-set implementation of time-dependent density functional theory for excited state dynamics simulations, J. Chem. Phys. **129**, 054110 (2008).

[30] P. Ordejón, E. Artacho, and J. M. Soler, Self-consistent order-N density-functional calculations for very large systems, Phys. Rev. B **53**, R10441 (1996).

[31] G. Kresse and J. Furthmüller, Efficient iterative schemes for ab initio total-energy calculations using a plane-wave basis set, Phys. Rev. B **54**, 11169 (1996).

[32] P. E. Blöchl, Projector augmented-wave method, Phys. Rev. B **50**, 17953 (1994).

[33] J. P. Perdew, J. A. Chevary, S. H. Vosko, K. A. Jackson, M. R. Pederson, D. J. Singh, and C. Fiolhais, Atoms, molecules, solids, and surfaces: Applications of the generalized gradient approximation for exchange and correlation, Phys. Rev. B **46**, 6671 (1992).

[34] B. Schonfeld, J. J. Huang, and S. C. Moss, Anisotropic mean-square displacements (MSD) in single-crystals of 2H- and 3R-MoS2, Acta Crystallogr., Sect. B **39**, 404 (1983).

[35] G. Henkelman, B. P. Uberuaga, and H. Jónsson, A climbing image nudged elastic band method for finding saddle points and minimum energy paths, J. Chem. Phys. **113**, 9901 (2000).

[36] G. Henkelman and H. Jónsson, Improved tangent estimate in the nudged elastic band method for finding minimum energy paths and saddle points, J. Chem. Phys. **113**, 9978 (2000).

[37] J.-U. Lee, S. Woo, J. Park, H. C. Park, Y.-W. Son, and H. Cheong, Strain-shear coupling in bilayer MoS2, Nat. Commun. **8**, 1370 (2017).

[38] R. T. Sam, T. Umakoshi, and P. Verma, Probing stacking configurations in a few layered MoS2 by low frequency Raman spectroscopy, Sci. Rep. **10**, 21227 (2020).

[39] J. Park, I. W. Yeu, G. Han, C. S. Hwang, and J.-H. Choi, Ferroelectric switching in bilayer 3R MoS2 via interlayer shear mode driven by nonlinear phononics, Sci. Rep. **9**, 14919 (2019).

[40] L. Liang, J. Zhang, B. G. Sumpter, Q.-H. Tan, P.-H. Tan, and V. Meunier, Low-Frequency Shear and Layer-Breathing Modes in Raman Scattering of Two-Dimensional Materials, ACS Nano **11**, 11777 (2017).

# Figures and Captions

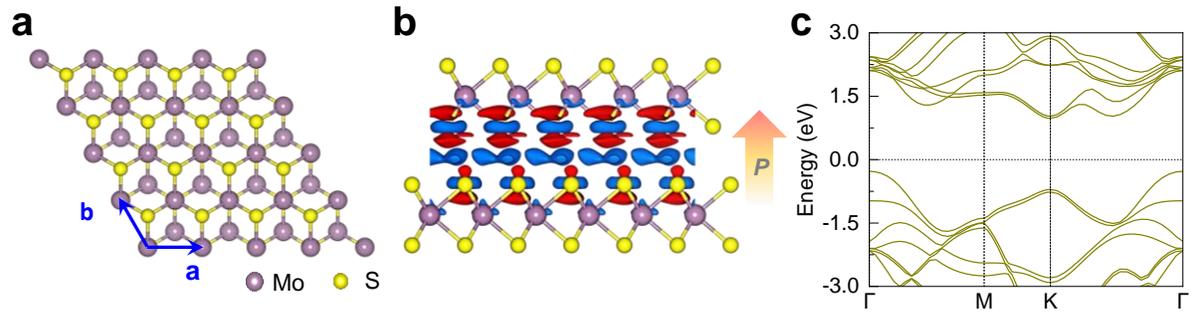

**Figure 1. Atomic and electronic structures of bilayer 3R-MoS$_2$.** (a) Crystal structure of bilayer 3R-MoS$_2$, with in-plane lattice vectors a and b shown in blue. (b) Charge density difference, plotted with an isosurface value of 0.0001 e/bohr$^3$. Red and blue colors denote regions of electron accumulation and depletion, respectively. (c) Band structure of bilayer 3R-MoS$_2$, with the Fermi level aligned to zero. The band structure calculations are performed at the level of the PBE functional.

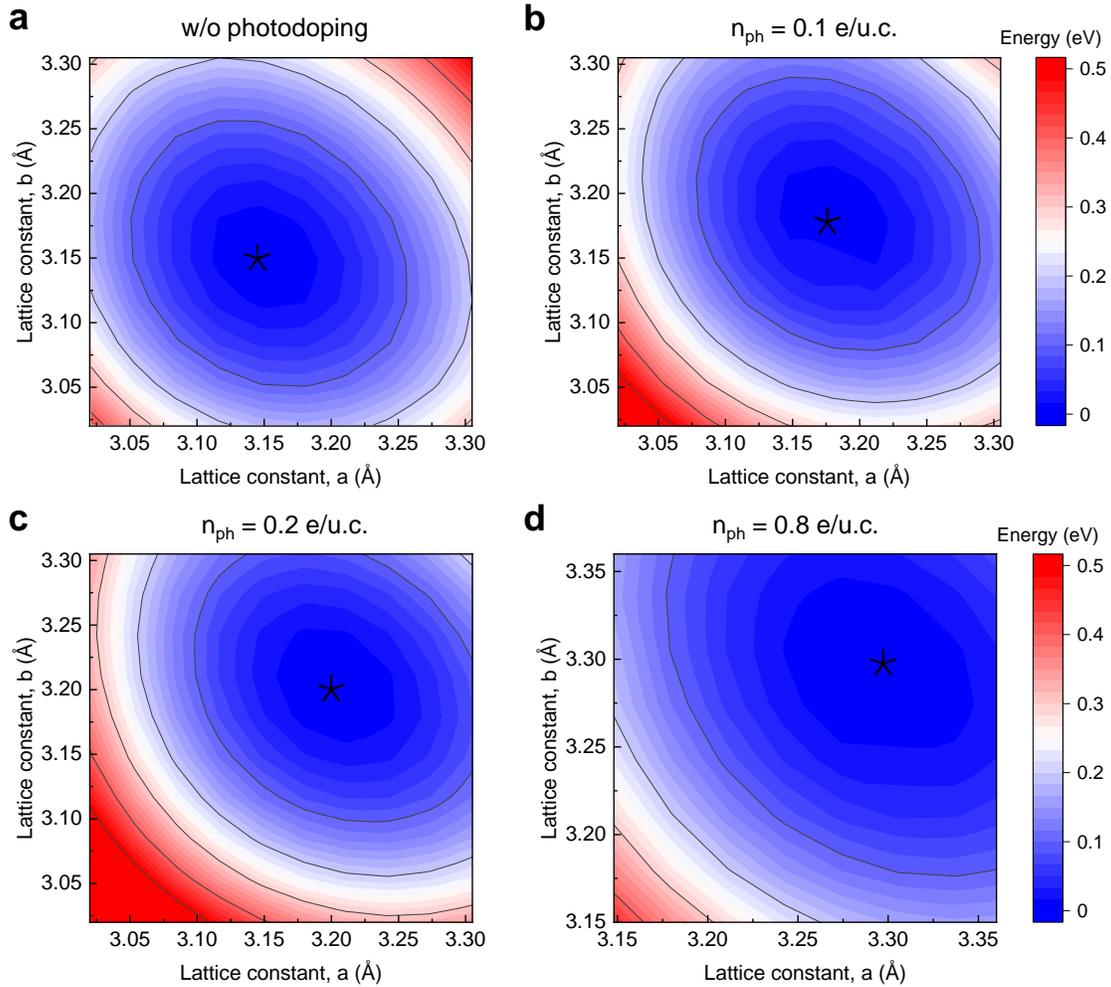

**Figure 2. Photodoping carrier-induced lattice expansion in bilayer 3R-MoS2.** (a) Energy diagram of the MoS$_2$ with different lattice constants without (w/o) photoexcitation. The energy minimum is located at a lattice constant of 3.15 Å. (b) The same quantity with (a) under the low photodoping level of 0.1 e/u.c.. The energy minimum is located at a lattice constant of 3.18 Å, with 1% expansion. (c) The same quantity with (a) under moderate photodoping level of 0.2 e/u.c.. The energy minimum is located at a lattice constant of 3.21 Å, with 2% expansion. (d) The same quantity with (a) under a high doping level (0.8 e/u.c.). The energy minimum is located at a constant of 3.30 Å, with 5% expansion. In all panels, **a** and **b** denote the lattice constants along the two directions. The energies are recalculated relative to the minimum in each panel. The stars denote he energy minima in each panel.

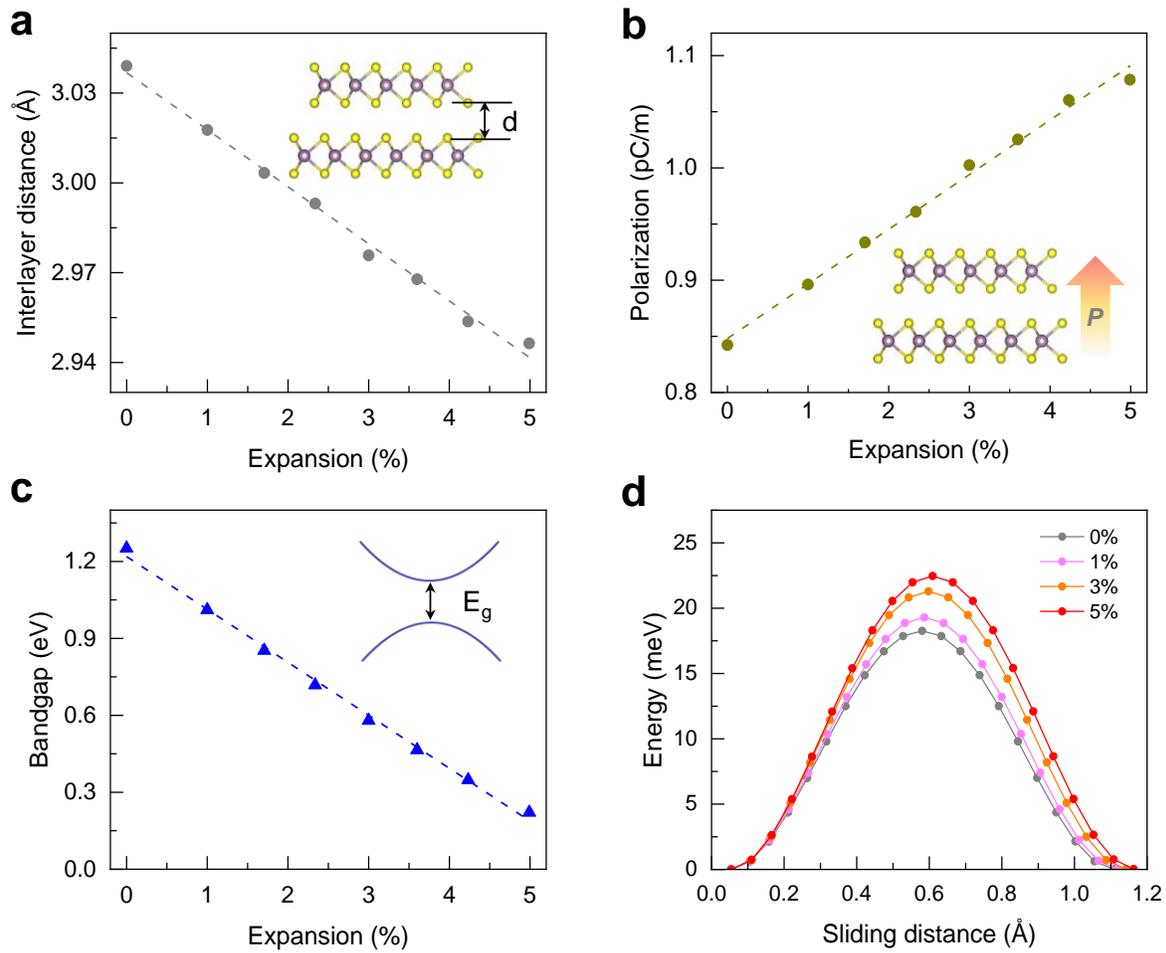

**Figure 3. Photostriction-tunable electronic properties in 3R-MoS₂.** (a) Interlayer distance with respect to lattice expansion, the inset illustrates the interlayer separation. (b) Polarization as a function of lattice expansion, the inset shows the direction of interlayer polarizations. (c) Bandgaps as a function of the lattice expansion. (d) Energy pathways for converting between two configurations (AA-up: 0 Å and AA-down: ~1.1 Å) under different expansions from 0% to 5%. The sliding distance represents the relative interlayer sliding distance between each transition state and the initial state along the transition pathway, which is interpolated by the nudged elastic band method.

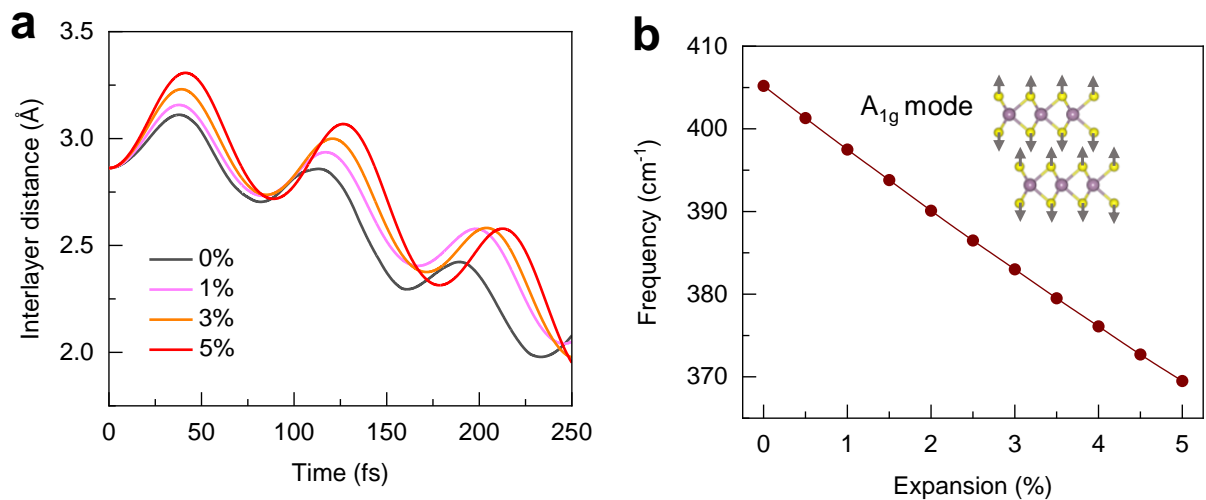

**Figure 4. Photostriction-tunable structural dynamics in 3R-MoS$_2$.** (a) Time-dependent interlayer distance with respect to lattice expansion from 1% to 5%, where the A$_{1g}$ mode dominates the lattice oscillation. The initial phonon mode is excited by compressing the interlayer distance along the out-of-plane direction. Subsequently, we propagate the electron–phonon coupled dynamics within the TDDFT framework and analyze the evolution of interlayer spacing following the excitation. (b) The frequency of the A$_{1g}$ mode with respect to lattice expansions, the inset illustrates the atomic displacements of the A$_{1g}$ mode.

# Supplemental Materials for
# "Photostriction-tunable Polarization and Structural Dynamics in Interlayer Sliding Ferroelectrics"

Kun Yang[1,2,#], Jianxin Yu[1,2,#], Jia Zhang[3], Sheng Meng[4,5], Jin Zhang[1,*]

[1]*Laboratory of Theoretical and Computational Nanoscience, National Center for Nanoscience and Technology, Chinese Academy of Sciences. Beijing 100190, China*

[2]*University of Chinese Academy of Sciences, Beijing 100049, China*

[3]*Max Born Institut für Nichtlineare Optik und Kurzzeitspektroskopie, Berlin 12489, Germany*

[4]*Beijing National Laboratory for Condensed Matter Physics, and Institute of Physics, Chinese Academy of Sciences, Beijing 100190, China*

[5]*Songshan Lake Materials Laboratory, Dongguan, Guangdong 523808, China*

*Corresponding author: Jin Zhang (jinzhang@nanoctr.cn)

**This file contains：**

Note. Regarding TDDFT and DFT methods.

S1. Distributions of photoexcited carriers in energy space.

S2. Charge density plots with different lattice expansions.

S3. Band structures of bilayer 3R-$MoS_2$ with different lattice constants.

## Note. Regarding TDDFT and DFT methods

**Time-dependent density functional theory.** The TDDFT calculations were performed utilizing the time-dependent *ab initio* package (TDAP), developed based on the time-dependent density functional theory [1-3] and implemented within SIESTA [4,5]. The dynamic simulations were carried out with an evolving time step of 50 as for both electrons and ions within a micro-canonical ensemble. To explain the TDDFT methods more, we present more description on our methods based on time-dependent Kohn-Sham equations for coupled electron-ion motion. We perform ab initio molecular dynamics for coupled electron-ion systems with the motion of ions following the Newtonian dynamics while electrons follow the time-dependent dynamics. The ionic velocities and positions are calculated with Verlet algorithm at each time step. When the initial conditions are chosen, the electronic subsystem may populate any state, ground or excited, and is coupled nonadiabatically with the motion of ions. We carried out k-space integration using a 15×15×1 mesh for the bilayer case in the Brillouin zone of the supercell to confirm the convergence in real-time propagation.

**Constrained density functional theory calculations.** This method modifies the electron density or potential to enforce specific electronic configurations, enabling the exploration of excited states, charge transfer, and other phenomena not naturally favored by standard DFT. It provides a systematic way to explore electronic structures that are not naturally favored by standard DFT but are relevant for understanding key physical and chemical properties in constrained environments [6].

The electron density [$\rho^{OCDFT}(r)$] in the general form of an orbital-constrained DFT expression is:

$$\rho^{OCDFT}(r) = \sum_i n_i^{OCDFT} \phi_i^*(r)\phi_i(r) \quad (1)$$

The total energy for the orbital-constrained DFT is written as $E_{KS}[\rho^{OCDFT}(r)]$ instead of ground-state $E_{KS}[\rho_0(r)]$. Herein $E_{KS}[\rho]$ is based on the orbital-constrained electron density, and $\rho_0(r)$ is the electron density in the ground state $\rho_0(r) = \sum_i n_i \phi_i^*(r)\phi_i(r)$; $n_i$ is the occupation number of the *i*-th orbital in ground state and $n_i^{OCDFT}$ is the fixed target occupation for the *i*-th orbital for the excited states.

To illustrate the TDDFT methods more, we present more description on our methods based on time-dependent Kohn-Sham equations for coupled electron-ion motion. We can perform ab initio molecular dynamics for coupled electron-ion systems with the motion of ions following the Newtonian dynamics while electrons following the TDKS dynamics. The ionic velocities and positions are calculated with Verlet algorithm at each time step. When the initial conditions are chosen, the electronic subsystem may populate any state, ground or excited, and is coupled nonadiabatically with the motion of ions.

This approximation works well for situations where a single path dominates in the reaction dynamics, for the initial stages of excited states before significant surface crossings take place, or for cases where the state-averaged behavior is of interest when many electron levels are involved as in condensed phases. However, the approach has limitations when the excited states involve multiple paths, especially when state specific ionic trajectories are of interest. In such cases, Ehrenfest dynamics fails as it describes the nuclear path by a single average point even when the nuclear wave function has broken up into many different parts. In this work, the deficiency is not critical because the laser induced path dominates in the whole reaction dynamics. In addition, we have tested the dynamics by altering the initial conditions such as ionic temperatures and laser intensities to confirm the robustness of our conclusion.

### Ground state calculations

The optimized atomistic and electronic structures are calculated by density functional theory (DFT) using the Vienna ab initio simulation package (VASP) [7]. The electronic wave function is described using projector-augmented wave (PAW) pseudopotential with the exchange correlation interaction treated using Perdew-Burke-Ernzerhof (PBE) method [8-10]. The van der Waals interactions between adjacent layers are described with optB86b-vdW functionals [11]. The plane wave cutoff energy is set as 550 eV, while the convergence thresholds for force and energy are 0.001 eV/Å and $10^6$ eV, respectively. A 21 × 21 × 1 k-grid sampling in the reciprocal space is adopted for layered 3R-$MoS_2$. Vacuum layers of approximately 15 Å are added to avoid interactions between adjacent slabs. The polarization properties are calculated using the Berry phase method [12].

## S1. Band structures of 3R-MoS$_2$ with different lattice constants.

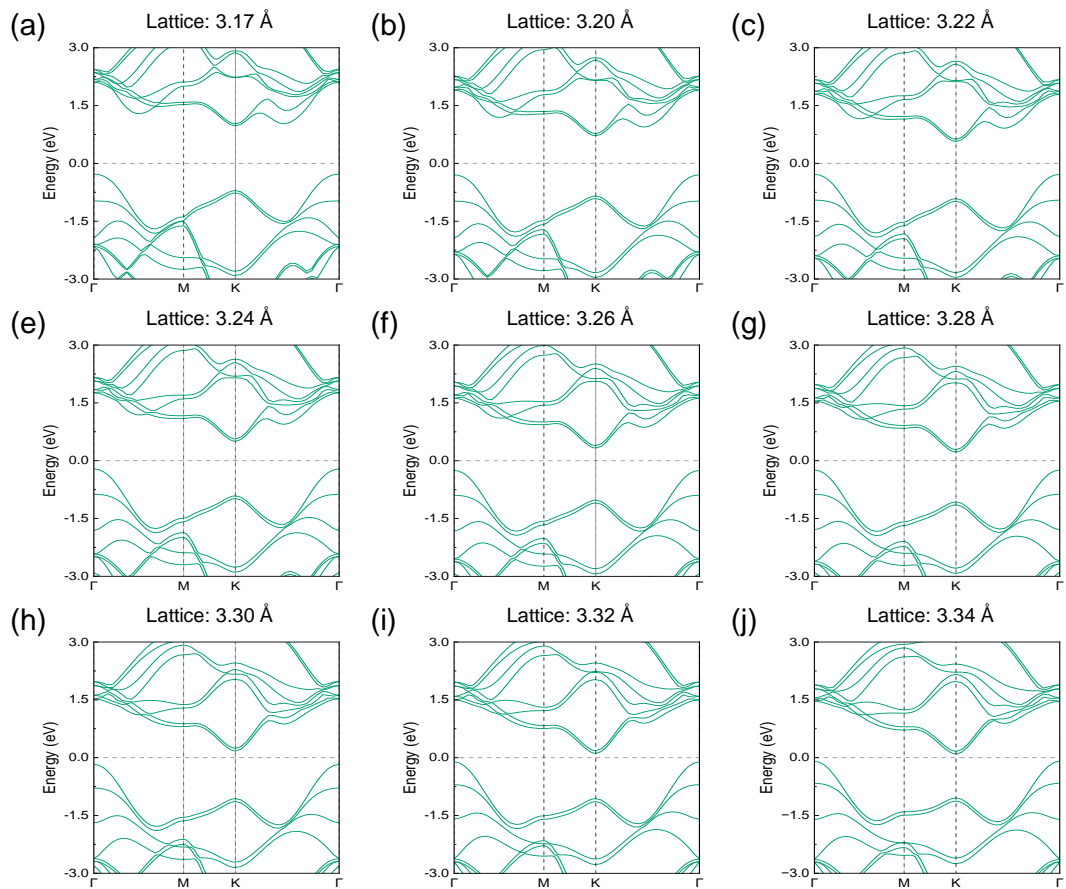

**Fig. S1.** Band structures of 3R-MoS$_2$ with different lattice constants. respectively. The band structures are calculated on the level of the PBE functional. The fermi levels are set as 0 eV.

## S2. Comparison between 3R an 2H stacked bilayer MoS$_2$

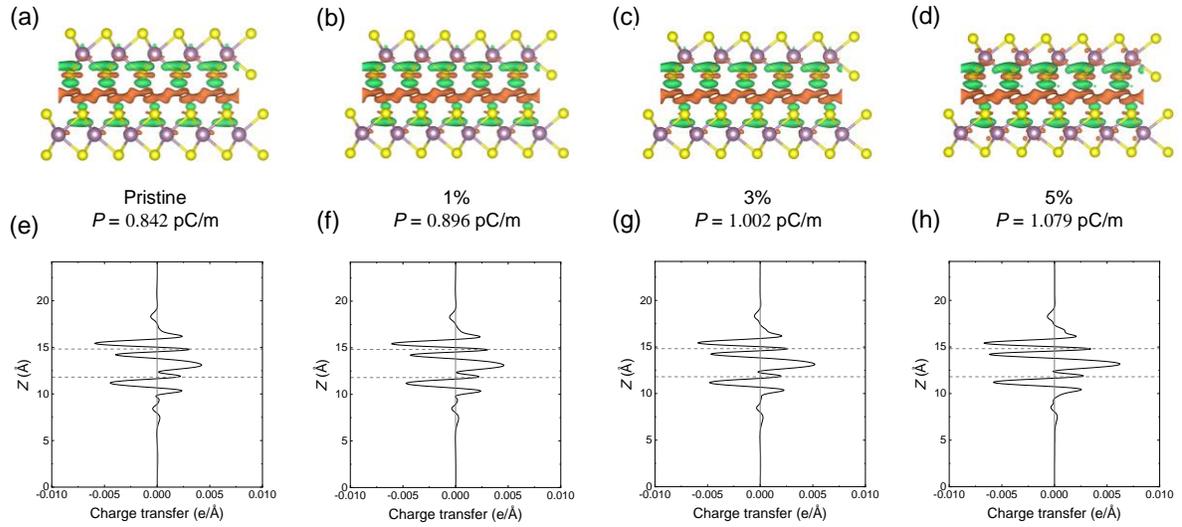

**Figure S2. Charge density plots.** (a-d) Interlayer differential charge density for different lattice expansions (0, 1%, 3% and 5%). The isosurface value of $1\times10^4$ e/bohr$^3$ was used for all panels. (e-h) The line profiles of charge difference along perpendicular directions, respectively.

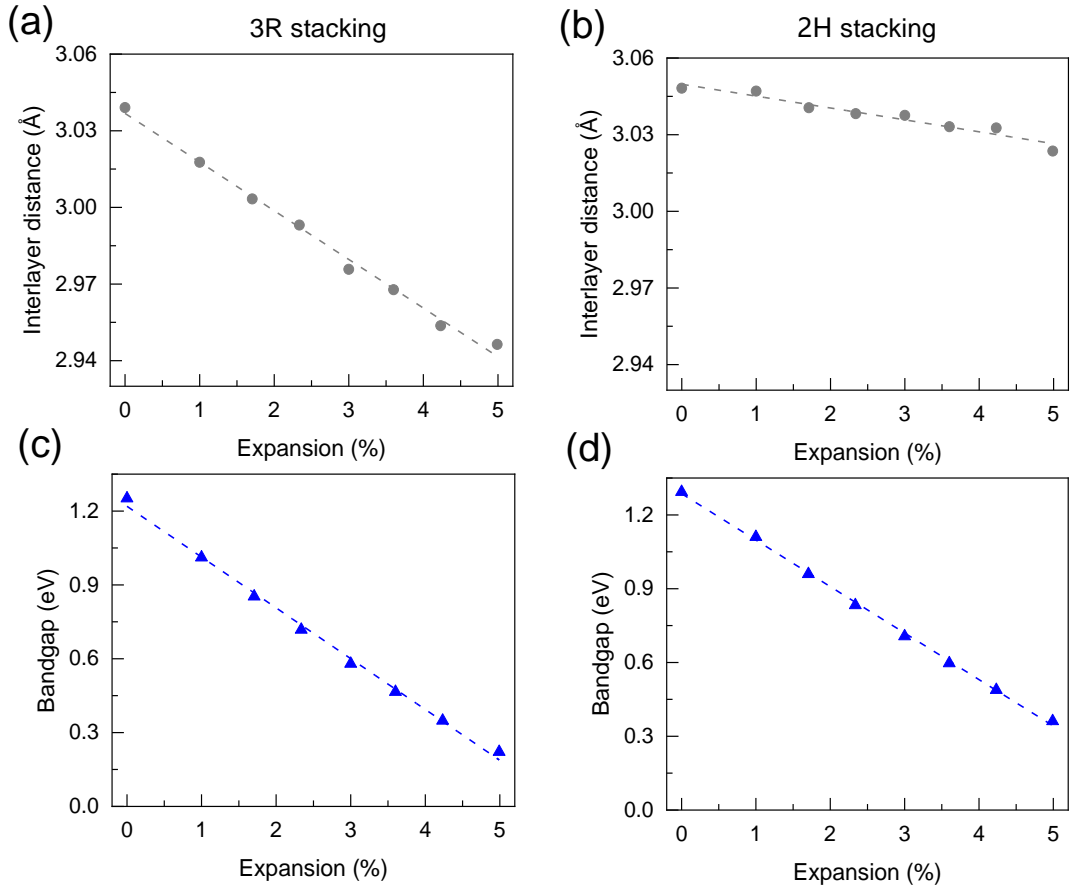

**Figure S3.** (a) Variation of the interlayer distance as a function of in-plane lattice expansion for different stacking configurations of MoS$_2$. The plot highlights the distinct interlayer relaxation behaviors in 3R and 2H stackings under tensile strain. (b) Evolution of the electronic bandgap with increasing lattice expansion, illustrating the stacking-dependent bandgap tunability. The 3R phase exhibits a more pronounced bandgap reduction compared to the 2H phase, reflecting stronger strain sensitivity of the electronic structure in non-centrosymmetric stacking.

## Reference

[1]  E. Runge and E. K. U. Gross, Density-Functional Theory for Time-Dependent Systems, Phys. Rev. Lett. **52**, 997 (1984).

[2] C. Lian, M. Guan, S. Hu, J. Zhang, and S. Meng, Photoexcitation in Solids: First-Principles Quantum Simulations by Real-Time TDDFT, Adv. Theory Simul. **1**, 1800055 (2018).

[3] S. Meng and E. Kaxiras, Real-time, local basis-set implementation of time-dependent density functional theory for excited state dynamics simulations, J. Chem. Phys. **129**, 054110 (2008).

[4] D. Sánchez-Portal, P. Ordejón, E. Artacho, and J. M. Soler, Density-functional method for very large systems with LCAO basis sets, Int. J. Quantum Chem. **65**, 453 (1997).

[5] M. S. José, A. Emilio, D. G. Julian, G. Alberto, J. Javier, O. Pablo, and S.-P. Daniel, The SIESTA method for ab initio order-N materials simulation, J. Phys.: Condens. Matter **14**, 2745 (2002).

[6] B. Kaduk, T. Kowalczyk, and T. Van Voorhis, Constrained Density Functional Theory, Chem. Rev. **112**, 321 (2012).

[7] G. Kresse and J. Hafner, Ab initio molecular dynamics for liquid metals, Phys. Rev. B **47**, 558 (1993).

[8] G. Kresse and J. Furthmüller, Efficient iterative schemes for ab initio total-energy calculations using a plane-wave basis set, Phys. Rev. B **54**, 11169 (1996).

[9] J. P. Perdew, J. A. Chevary, S. H. Vosko, K. A. Jackson, M. R. Pederson, D. J. Singh, and C. Fiolhais, Atoms, molecules, solids, and surfaces: Applications of the generalized gradient approximation for exchange and correlation, Phys. Rev. B **46**, 6671 (1992).

[10] P. E. Blöchl, Projector augmented-wave method, Phys. Rev. B **50**, 17953 (1994).

[11] J. Klimeš, D. R. Bowler, and A. Michaelides, Phys. Rev. B **83**, 195131 (2011).

[12] R. D. King-Smith and D. Vanderbilt, Theory of polarization of crystalline solids, Phys. Rev. B **47**, 1651 (1993).